\documentclass{jnmp01b}

\usepackage{amsmath}

\numberwithin{equation}{section}

\newcommand {\fwitt}  {{\mathfrak{witt}}}
\newcommand {\Cee}    {{\mathbb  C}}
\newcommand {\pder}[1] {{\frac{\partial}{\partial {#1}}}}
\newcommand {\fg}     {{\mathfrak{g}}}    %
\newcommand {\fosp}   {{\mathfrak{osp}}}
\newcommand {\fgl}    {{\mathfrak{gl}}}  %
\newcommand {\fsl}    {{\mathfrak{sl}}}
\newcommand {\fvir}   {{\mathfrak{vir}}}

\begin{document}

\renewcommand{\evenhead}{D.\ Leites}
\renewcommand{\oddhead}{How to Superize Liouville Equation}


\thispagestyle{empty}

\begin{flushleft}
\footnotesize \sf
Journal of Nonlinear Mathematical Physics \qquad 2000, V.7, N~3,
\pageref{firstpage}--\pageref{lastpage}.
\hfill {\sc Letter}
\end{flushleft}

\vspace{-5mm}

\copyrightnote{2000}{D.\ Leites}

\Name{How to Superize Liouville Equation}

\label{firstpage}

\Author{Dimitry LEITES}

\Adress{Department of Mathematics, University of Stockholm, Roslagsv.\ 101, \\
Kr\"aftriket hus 6, 
S-106 91, Stockholm, Sweden\\ 
E-mail: mleites@matematik.su.se}

\Date{Received September 18, 1999; Revised March 1, 2000; 
Accepted April 1, 2000}

\begin{abstract}
\noindent
So far, there are described in the literature two
ways to superize the Liouville equation: for a scalar field (for
$N\leq 4$) and for a vector-valued field (analogs of the
Leznov--Saveliev equations) for $N=1$.  Both superizations are
performed with the help of Neveu--Schwarz superalgebra.  We consider
another version of these superLiouville equations based on the Ramond
superalgebra, their explicit solutions are given by Ivanov--Krivonos'
scheme. Open problems are offered.
\end{abstract}



\section{A missed superization of the Liouville equation}
The Liouville equation 
\begin{equation*}
\frac{\partial^2}{\partial x\partial y} f=\exp(2f)\tag{1.1}
\end{equation*}
so important in applications ranging from SUGRA to soap manufacturing
(see, e.g., \cite{P}, \cite{GSW}, \cite{Pe}), \cite{LeSa} is manifestly invariant
with respect to the Witt algebra $\fwitt$, the Lie algebra of
conformal transformations:
\begin{equation*}
x\mapsto a(x), \; y\mapsto b(y), \; f\mapsto f+\ln a+\ln b\text{ for 
}x, y\in\Cee.   \tag{1.2}
\end{equation*}
This invariance provides us with a ``general'' (in the sense that it 
belongs to an open orbit) solution
\begin{equation*}
\exp(f)=\frac{a_x{}'b_y{}'}{(1-ab)^2}.\tag{1.3}
\end{equation*}
For super version of (1.1) one invariably takes (\cite{P}, 
\cite{GSW})
\begin{equation*}
D_+D_-f=m\exp(f), \text{ where }D_\pm =\pder{\theta_\pm}+ 
\theta_\pm\pder{x_\pm}.\tag{1.4}
\end{equation*}
Equation (1.4) was first integrated in \cite{LLS} in components. The 
solution in terms of superfields was obtained by Ivanov and Krivonos 
\cite{IK}:
\begin{equation*}
\exp(-f)=\displaystyle\frac{D_+(\frac{D_+a}{\sqrt{D_+^2(a)}})D_-(\frac{D_-b}
{\sqrt{D_-^2(b)}})}{m^2a-b+m\frac{D_-bD_+a}{\sqrt{D_-^2(b)D_+^2(a)}}}.
\tag{1.5}
\end{equation*}
A completely integrable version of (1.4) for a vector-valued function 
$f=(f_1, \dots f_n)$ (together with a way to get an explicit solution) 
was given in \cite{LSS} for the systems (equivalent for invertible 
matrix $A$):
\begin{equation*}
D_+D_-f_i=\exp(\mathop{\sum}\limits_{j}A_i^jf_j) \text{ and } 
D_+D_-f_i=\mathop{\sum}\limits_{j}A_i^j\exp(f_j) , 1\leq i\leq n 
\tag{1.6}
\end{equation*}
where $A$ is the Cartan matrix of the Lie superalgebra $\fg(A)$ 
admitting a superprincipal embedding of $\fosp(1|2)$ (all such 
embeddings are classified in \cite{LSS}).  Simultaneously, \cite{A}, 
see also \cite{PZ}, observed that a wider class of Cartan matrices is 
worth considering, namely, all algebras $\fg(A)$ of polynomial growth 
with a system of simple roots all of which are odd.

The aim of this note is to draw attention to ``missed opportunities'' 
in superization of (1.1).  Namely, in (1.4)--(1.6) and in the solution 
of (1.6) we may replace the above $D_\pm$ (which are copies of the 
contact field $K_{\theta_{\pm}}$ described in \cite{GLS} and 
corresponding to the Neveu-Schwarz algebra) with
\begin{equation*}
\tilde D_\pm =\frac12\tilde K_{\theta_{\pm}}=\theta_{\pm}
\pder{t_{\pm}}-\frac{1}{t_{\pm}}\pder{\theta_{\pm}}\tag{1.7}
\end{equation*}
which correspond to the Ramond superalgebra.  Observe that since the 
algebras of invariance of the equations corresponding to 
Neveu--Schwarz and Ramond superalgebras are non-isomorphic, these 
equations are essentially different though their underlying equations 
in Bose sector are equivalent (accordingly, the even parts of NS and 
R are isomorphic).

The above applies also to the superizations of the Brockett equation 
\cite{Sa} and Saveliev-Vershik's version \cite{SV} of the 
Liouville equation
\begin{equation*}
\frac{\partial^2}{\partial x\partial y} f=K \exp(f)\tag{1.8}
\end{equation*}
associated with the {\it continuum} Lie superalgebras, cf.  \cite{SV}, 
i.e., when the non-existent Cartan matrix is replaced with a 
{\it nonlinear} operator $K$.  Recently B.~Shoikhet and A.~Vershik 
\cite{ShV} found an explicit form of $K$ for 
$\fgl(\lambda)=L(U(\fsl(2))/(\Omega-\lambda^2+1))$, where $\Omega$ is 
the quadratic Casimir operator of $\fsl(2)$ and $L(A)$ is the Lie 
algebra constructed on the space of the associative algebra $A$ by 
replacing the dot product with the bracket, cf.  also \cite{GL2} and 
\cite{V}, where more general Lie (super)algebras based on any simple 
$\fg$, rather than $\fsl(2)$, are considered.  

\section{Open problems}

1) There are 4 series and several exceptional simple stringy 
(``superconformal'') Lie superalgebras that generalize $\fwitt$, see 
\cite{GLS} (or \cite{CK}).  All are realized on the supercircle of 
dimension $1|N$.  For Lie superalgebras with polynomial coefficients 
corresponding to some stringy superalgebras with $N\leq 4$ Ivanov and 
Krivonos suggested a scheme for explicit integration of the analogs of 
the Liouville equation for the {\it scalar} field and executed it for 
small values of $N$.  They did not consider the higher $N$ for two 
reasons: (a) the volume of calculations is too high to undertake 
without serious motivations, (b) for $N>4$ spin $>2$.  In view of 
Vasiliev's ideas \cite{V} (b) can be considered as an obsolete and 
non-existing obstacle, and in view of a new technique \cite{GL1} the 
volume of calculations might become tolerable.

What does Ivanov--Krivonos' scheme give for greater $N$ and 
other stringy superalgebras?

2) How to apply Ivanov--Krivonos' scheme for $N>1$ 
to the matrix equations from \cite{LSS} (or the other way round)? This 
blend should be associated with a generalization of the 
``superprincipal embedding''.  Though its meaning for $N>1$ is unclear; an 
approach is suggested in \cite{GL2}.

3) The 12 of the simple stringy Lie superalgebras are {\it 
distinguished}: only they have nontrivial central extensions.  
(Observe that since one of the distinguished superalgebras has 3 
nontrivial central extensions each, there are exactly 14 direct 
superizations of the Schr\"odinger equation, KdV hierarchies and 
related structures, see \cite{LX}.) 

Ivanov-Krivonos's scheme seem to require simple Lie superalgebras
$G$ of vector fields with polynomial coefficients, not Laurent ones
$G^L$.  Unlike $G^L$, algebras $G$ have no central extension, so the
importance of the distinguished stringy superalgebras for integration
of the analogs of Liouville equation is unclear.  Recently, however,
F.~Toppan demonstrated \cite{T} (unpublished) that the central charge
of the Virasoro algebra $\fvir$, the central extension of $\fwitt$,
non-trivially acts on (1.1).  This miraculous result indicates that
probably there are other, inner, mechanisms that restrict application
of Ivanov-Krivonos's scheme for non-distinguished  stringy Lie
superalgebras $G^L$.  What are these
mechanisms, if any?

4) I considered the above equations over complex numbers.  In
applications functions of real variable are often preferable.  This
leads to necessity to classify real forms of stringy Lie
superalgebras.  Though this is done for almost all simple stringy
superalgebras, see \cite{S1} (and the results are vital even for the
description of real forms of Kac--Moody algebras, cf.  \cite{S2}), the
answer seems to be unknown to physicists being published in a purely
mathematical journal.  The description of the real forms of the
remaining simple stringy superalgebras is in preparation; to describe
all real solutions for higher $N$ and single out physically relevant
ones is a problem, cf.  \cite{IK}.

5) (Due to a referee.)  One may expect, on purely aesthetic
grounds, that in super case there {\bf must} exist some kind of
superstructure over the {\bf set} (1.5) of solutions of super
Liouville equations.  I.e., on a deeper level there must be a
``superset'' of solutions on which an infinite dimensional supergroup
({\sl not} Lie!)  of automorphisms acts, so that the corresponding
``superset'' is realized as an open superorbit of this
supergroup.

To extend the constructs of Ivanov-Krivonos scheme to this level as
well, one needs to have, evidently, the super version of
Kac--Peterson's theory (cf.  \cite{K}) of groups associated with
integrable algebras.

\newpage

\subsection*{Acknowledgements}

Financial support of NFR and illuminating discussions with 
E.~Ivanov, S.~Krivonos, B.~Shoikhet, F.~Toppan and A.~Vershik are 
gratefully acknowledged.

\label{lastpage}


\begin{thebibliography}{99}
\small

\bibitem{A}
Andreev V.A., Supersymmetric Generalized Toda Lattice (Russian), in 
Group-Theoretic Methods in Physics, Vol.\ 1 (Russian), 
Editors M.\ Markov et.\ al.,  
Yurmala, 1985, Nauka, Moscow, 1986, 439--452, 
English translation: VNU Sci.  Press, Utrecht, 1986, 315--321; 
Odd Bases of Lie Superalgebras and Integrable Systems, {\em Theor.  Math.  
Phys.},  1987, V.72, 758--764.

\bibitem{CK}
Cheng S.-J.\ and Kac V., A New $N = 6$ Superconformal Algebra,  
{\em Commun.  Math.  Phys.}, 1997, V.186, N~1, 219--231.

\bibitem{GL1}
Grozman P.\ and Leites D., {\it Mathematica}-Aided Study of Lie 
Algebras and Their Cohomology.  From Supergravity to Ball Bearings and 
Magnetic Hydrodynamics, in  The second 
International Mathematica symposium, Editor V.\ Ker\"anen, 
Rovaniemi, 1997, 185--192.

\bibitem{GL2}
Grozman P.\ and Leites D., Lie Superalgebras of Supermatrices of Complex 
Size.  Their Generalizations and Related Integrable Systems,
In Complex Analysis and Related Topics,
N.\ Vasilevsky et.\ al.,
Proc.\ Internatnl.\ Symp.\ Mexico, 1996, Birkhauser Verlag, 
1999, 73--105.

\bibitem{GLS} 
Grozman P., Leites D.\ and Shchepochkina I., Lie Superalgebras of String 
Theories, hep-th/9702120.

\bibitem{GSW}
Green M., Schwarz J.\ and Witten E. Superstring theory, vv.1, 2, 
Cambridge Univ.  Press, Cambridge, 1987.

\bibitem{IK}
Ivanov E.A.\ and Krivonos S.O., Integrable Systems as Nonlinear 
Realizations of Infinite-Dimensional Symmetries: The Liouville 
Equation Example,  {\em Lett.  Math.  Phys.},  1984, V.8, N~1, 39--45; 
${\rm U}(1)$-Supersymmetric Extension of the Liouville Equation,  
{\em Lett.  Math.  Phys.}, 1983, V.7, N~6, 523--531,  
Errata: ``${\rm U}(1)$-Supersymmetric Extension of the Liouville Equation''.  
{\em Lett.  Math.  Phys.}, 1984,  V.8, N~4, 345; 
Nonlinear Realization of the 
Conformal Group in two Dimensions and the Liouville Equation 
(Russian), {\em Teoret.  Mat.  Fiz.}, 1984, V.58, N~2, 200--212; 
Superfield Extensions of the Liouville Equation (Russian), 
in Proceedings of the VII International Conference on the Problems of 
Quantum Field Theory (Russian), Alushta, 1984, 257--271, JINR, Dubna, 
D2-84-366, JINR, Dubna, 1984; 
$N=4$ Super-Liouville Equation,  
{\em J. Phys.  A}, 1984, V.17, N~12, L671--L676; 
id., 
$N=4$ Superextension of 
the Liouville Equation with a Quaternion Structure  (Russian), 
{\em Teoret.  Mat.  Fiz.}, 1985,  V.63, N~2, 230--243.

\bibitem{K}
Kac V.G., Infinite Dimensional Lie Algebras,  3rd ed,  Cambridge 
Univ.  Press, Cambridge, 1992.

\bibitem{LLS}
Leznov A.N., Leites D.A.\ and Saveliev M.V., Superalgebra $B(0, 1)$ and 
Explicit Integration of the Supersymmetric Liouville Equation.  {\em ZhETF 
Letters}, 1980, V.32, N~1, {\em Physics Letters}, 1980. V.96B, 97--99.

\bibitem{LSS} 
Leites D.A., Saveliev M.V. and Serganova V.V., Embeddings of the Lie 
Superalgebra $\fosp(1|2)$ and the Nonlinear Supersymmetric Equations 
Associated with Them  (Russian),  
Group-Theoretic Methods in Physics, Vol.  1 (Russian),
Editors Markov M. et.  al.,
Yurmala, 1985, Nauka, Moscow, 1986, 377--394, 
English translation: VNU Sci. Press, Utrecht, 1986, 255--297.

\bibitem{LX} 
Leites D.\ and Xuan P., Supersymmetry of the Schr\"odinger and 
Korteweg--de Vries Operators, hep-th9710045.

\bibitem{LeSa} 
Leznov A.\ and Saveliev M., Group-Theoretical Methods for Integration
of Non-Linear Dynamical Systems (Russian), Moscow, Nauka, 1985,
English version: Translated and revised from the Russian,  Translated
by D.A. Leites,  Progress in Physics, 15,  Birkh\"auser Verlag,
Basel, 1992,  xviii+290 pp.

\bibitem{PZ}
Penati S.\ and Zanon D., Supersymmetric, Integrable Toda Field Theories: 
The B(1,1) Model, {\em Phys.  Lett.  B}, 1992, V.288, 297--305.
 
\bibitem{Pe} 
Peliti L., Amphiphilic Membranes, cond-mat/9501076, (92 pages, 43 
figures; Lectures presented at the 1994 Les Houches Summer School 
``Fluctuating Geometries in Statistical Mechanics and Field Theory.'')

\bibitem{P} 
Polyakov A.M., Quantum Geometry of Bosonic Strings,  {\em Phys.  Lett. B},
1981, V.103, 207--210; 
Quantum Geometry of Fermionic
Strings,  {\em Phys.  Lett. B}, 1981, V.103, 211--213.

\bibitem{ShV} 
Shoikhet B.\ and Vershik A., Lie Algebras with Continual Root System and 
Cartan Subalgebra is $\Cee[h]$,  
In  Proc.  Internat.  Seminar on integrable systems (In memoriam M.~Saveliev),
Editors Manin Yu.\  et al., 
(www.mpim-bonn.mpg.de), MPI-1999-36, 1999, 137--140.

\bibitem{SV} 
Saveliev M.\ and Vershik A., Continuum Analogs of Contragredient Lie 
Algebras, {\em Commun. Math. Phys.}, 1989, V.126, 367--378.

\bibitem{Sa} 
Saveliev M., On Specializing the Brockett Equation for Nonabelian 
Versions of the Generalized Toda Lattices,  {\em Phys.  Lett.  A}, 1991,
V.160, N~4, 355--356.

\bibitem{S1}
Serganova V., Outer Automorphisms and Real Forms of Kac--Moody 
Superalgebras  (Russian),  
In  Group theoretical methods in physics, Vol.  I, Proc.  Int.  Semin.
(Russian),
Editors M.\ Markov et.  al.,
Zvenigorod, 1982, 
Nauka, Moscow, 1983, 279--283.

\bibitem{S2}
Serganova V., Automorphisms of Lie Superalgebras of String Theories  
(Russian), {\em Funktsional.  Anal.  i Prilozhen.}, 1985,  V.19, N~3, 75--76, 
(a detailed version in: Seminar on Supermanifolds, 
Editor D. Leites, 
Reports of Stockholm University, 1987--1992, v.  
22/1988-4).

\bibitem{T}
Toppan F., On Anomalies in {\em Classical} Dynamical
Systems, in preparation.

\bibitem{V}
Vasiliev M.A., Higher-Spin Gauge Theories in Four, Three and Two 
Dimensions,  The Sixth Moscow Quantum Gravity Seminar, 1995, 
{\em Internat.  J. Modern Phys.  D}, 1996,  V.5, N~6, 763--797; 
Higher-Spin-Matter Gauge Interactions in $2+1$ Dimensions.  Theory of 
Elementary Particles (Buckow, 1996), Nuclear Phys.  B Proc.  Suppl.  
,56B, 1997, 241--252.

\end{thebibliography}
\end{document}